\documentclass[12pt]{article}

\usepackage[utf8]{inputenc}
\usepackage{amssymb}
\usepackage{amsmath}
\usepackage{subcaption}
\usepackage{graphicx}
\usepackage[margin=2.3cm]{geometry}
\usepackage{natbib}
\usepackage{xcolor}
\usepackage[official]{eurosym}
\usepackage{wrapfig}
\usepackage{booktabs}
\usepackage{graphics}
\usepackage[bookmarks=false]{hyperref}
\usepackage[graphicx]{realboxes}
\usepackage{bm}
\numberwithin{equation}{section}
\usepackage{siunitx}
\usepackage{gensymb}
\usepackage{makecell}
\usepackage[T1]{fontenc}
\usepackage{titling}
\usepackage{authblk}
\usepackage{appendix}
\usepackage{enumitem}   

\title{Accounting for climate change in extreme sea level estimation}
\author{Eleanor D'Arcy$^*$}
\author{Jonathan A. Tawn}
\affil{\small STOR-i Centre for Doctoral Training, Department of Mathematics and Statistics, Lancaster University, LA1 4YR, UK \\ \footnotesize{$^*$ Correspondence to: e.darcy@lancaster.ac.uk}}
\date{\today}

\begin{document}

\maketitle

\begin{abstract}
 Extreme sea level estimates are fundamental for mitigating against coastal flooding as they provide insight for defence engineering. As the global climate changes, rising sea levels combined with increases in storm intensity and frequency pose an increasing risk to coastline communities. We present a new method for estimating extreme sea levels that accounts for the effects of climate change on extreme events that are not accounted for by mean sea level trends. We follow a joint probabilities methodology, considering skew surge and peak tides as the only components of sea levels. We model extreme skew surges using a non-stationary generalised Pareto distribution (GPD) with covariates accounting for climate change, seasonality and skew surge-peak tide interaction. We develop methods to efficiently test for extreme skew surge trends across different coastlines and seasons. We illustrate our methods using data from four UK tide gauges.
\end{abstract}


\newpage


\section{Introduction}\label{intro}
\begin{figure}
    \centering
    \includegraphics[width=0.5\textwidth]{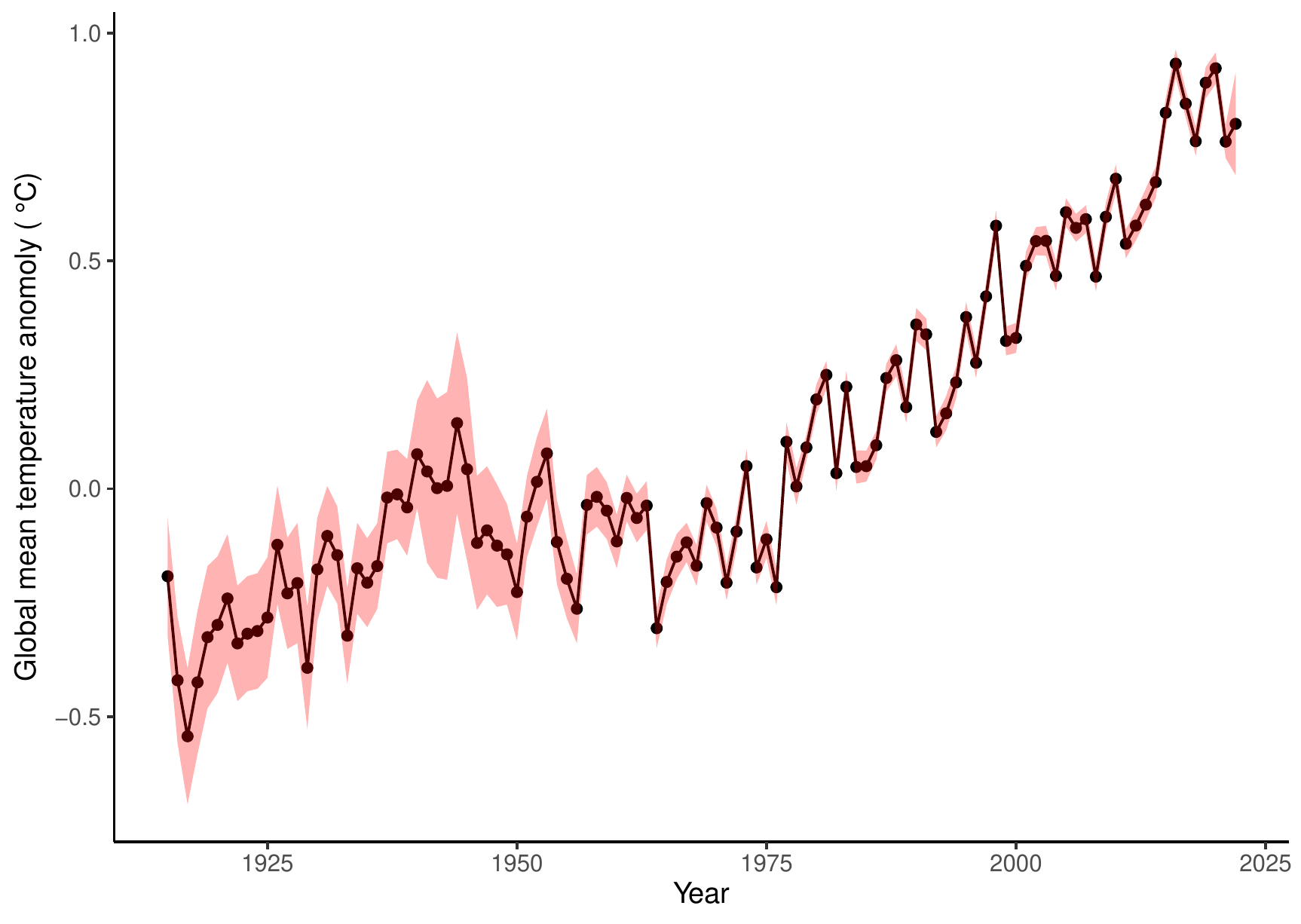}
    \caption{Global mean temperature anomalies from 1915 to 2020, relative to the period 1961-1990, with associated uncertainties in red.}
    \label{fig::GMT}
\end{figure}

The UK coastline is one of the largest in Europe at approximately 8000km for mainland Britain and is regularly subject to coastal flooding~\citep{zsamboky2011}. Coastal flooding is defined as a natural phenomenon where coastal land is inundated by sea levels above the normal tidal conditions. This has the potential to devastate coastal towns, damage infrastructure and destroy habitats. In extreme cases, coastal flooding has led to the loss of human life. The likelihood of coastal flooding is increasing with anthropogenic induced climate change (see Figure~\ref{fig::GMT}) so it is increasingly important to protect coastline communities or at least have a well-founded scientific basis for the proposal for a managed retreat. Coastal flood defences, such as a sea wall, protect against these consequences if they are built to withstand the most extreme sea levels. However, resources are wasted in building defences that are too conservative. Estimates of sea level return levels provide crucial information for this design process; a return level is the value we expect the annual maximum sea level to exceed with probability $p$, i.e., once every $1/p$ years, on average, for a stationary series. We estimate these levels for $p\in[10^{-4},10^{-1}]$ to cover a range of industry interest, from agricultural preservation to nuclear power plant protection. 

Coastal flooding is influenced by a combination of tide, surge and waves. We are interested in the still water level, i.e., the sea level with waves filtered out, but for simplicity we refer to this as sea level. Therefore, tide and surge are the only components of sea levels that we consider. Tides are the regular and predictable changes in sea levels driven astronomically; these changes are well understood and perfectly forecast~\citep{egbert2017}. High tides generally occur once every 12 hours and 25 minutes, although variations are possible. We refer to the maximum tide in this cycle as the peak tide. Surges are stochastic, transient changes in sea levels often caused by strong winds and low atmospheric pressure due to a storm, hence are often referred to as storm surges.  Surges are sometimes called the non-tidal residual as they define any departure from the predicted tidal regime so can also include gauge recording errors, tidal prediction errors and effects of the tide-surge interaction. These are often available at hourly or 15 minute intervals on the UK National Tide Gauge Network. We refer the reader to \cite{pughwoodworth2014} for a complete overview of sea level processes. 

An alternative decomposition of sea levels is to consider the maximum level in a tidal cycle that can be written as the sum of skew surge and peak tide. Skew surge is the difference between the maximum observed sea level and the peak tide in a tidal cycle, regardless of their timing. In this case, we have less data since observations are available once every tidal cycle. However, skew surge and peak tide exhibit a much weaker dependence than surge and tide (which has a complex dependence structure), so they are often preferred.~\cite{Williams2016} show that it is reasonable to assume skew surge and peak tide are independent at most sites on the UK National Tide Gauge Network; though there is physical evidence that this is not always true~\citep{howardWilliams2021}.

Long term changes in mean sea level have been widely studied~\citep{woodworth2003, wahl2013} via empirical assessments and using hydrodynamic models linked to climate models. Typically linear models are fit to estimate these trends. Similar statistical methods have been used for extreme sea levels using regression of annual or monthly maximum data on either sea levels or skew surges. Interestingly these simple exploratory methods find no significant evidence for the trend in extreme sea levels to differ from that for mean sea levels~\citep{wong22,woodworth2011,weisse2014}. Complications to these methods are the large interannual variability, the presence of seasonality and the inefficient usage of extreme event data (through the use of maxima rather than all large values). The difference between extreme and mean sea level trends is likely to be of a smaller order than for mean sea level trend, hence they are more difficult to estimate. Furthermore, only trends in average extreme values are looked for, not changes in their variability over time. As a consequence, inference for these properties at a single site is likely to be overloaded by uncertainty, resulting in the hypothesis of identical trends in extreme and mean sea levels not being rejected. 

We propose a different approach which is integrated into the estimation of return levels for extreme sea levels; this accounts for short term variations in skew surges such as seasonality, uses all extreme skew surges above a high month specific quantile, allows for the distribution of the extreme skew surges, and enables pooling of information about the trend across sites. Critically, we separately assess changes in the rate of which extreme skew surges events occur and changes in the distribution (e.g., the mean) of these extreme events once they have occurred.

The earliest methods for estimation of sea level return levels modelled the sea level data directly, whilst the next approaches used a joint probabilities method to consider surge and tide components. More recent approaches use skew surge and peak tides. Section~\ref{sec::ex_methods} gives an overview of the history of methodology developments. We extend the most recent method, that of~\cite{DArcy22}, to account for the effects of climate change on extreme sea level estimation. They model skew surges and combine this with the known peak tide regime. They particularly focus on the tail of the skew surge distribution, using a generalised Pareto distribution (GPD) to model exceedances of a high threshold~\citep{Coles2001}. Covariates are added to this model representing day of the year and peak tide, to account for seasonality and skew surge-peak tide dependence, as well as capturing the temporal dependence of extreme skew surges. Their results demonstrate a considerable improvement on previous approaches since the realism of the sea level processes is captured, and significant improvements in goodness-of-fit are achieved. However, their model assumed that skew surges were identically distributed across years, after a linear mean sea level trend was removed. If climate change impacted the within year skew surge variance, or even its distribution in a more subtle way than simply changing its mean value, then the extreme sea levels relative to the mean sea level will also change. Therefore we need a methodology that can incorporate such changes through to the return level estimation. Here we will develop methods to account for non-stationarity in this mean adjusted skew surge data to help quantify any remaining non-stationarity in extreme skew surges.

In Sections~\ref{sec::data} and~\ref{sec::EVA} we introduce the data and relevant extreme value theory, respectively. Section~\ref{sec::ex_methods} reviews the exisiting methods used for extreme sea level estimation, with a particular focus on that from~\cite{DArcy22}. In Section~\ref{sec::trends} we propose methods for investigating longer-term trends in extreme skew surges, with respect to time and global mean temperature anomaly (GMT), at a single site. We consider pooling information about the trends in extreme value model across sites in Section~\ref{sec::dependence}, and suggest methods for exploring pairwise extremal dependence in skew surges across sites. We present the results for the single site and pooled methods in Sections~\ref{sec::trends_results} and~\ref{sec::pooling_results}, respectively. Section~\ref{sec::discussion} concludes this paper with a summary of our findings and suggestions for future work.

\section{Materials and Methods}

\subsection{Data}\label{sec::data}
Sea level observations are taken from the UK National Tide Gauge Network maintained by the Environment Agency. The data undergo rigorous quality control and can be obtained from the British Oceanographic Data Centre (BODC). This network is part of the National Tidal and Sea Level Facility where tidal elevations are recorded at 44 sites along the UK coastline. We consider data from Heysham, Lowestoft, Newlyn and Sheerness, located on the west, east, south and east (at the Thames Estuary) coast of England, respectively. Newlyn has the longest observational period from 1915-2020 (17\% missing), whereas Heysham, Lowestoft and Sheerness have data available for 1964-2016 (17\% missing), 1964-2020 (4\% missing) and 1980-2016 (9\% missing), respectively. Observations are given as a height in metres above chart datum.

We chose to study these sites because they have different characteristics, are typically affected by different storms and all have a long observational periods. Heysham has the second largest tidal range on the network and is a tidally dominant site, whereas Lowestoft is surge dominant. Sheerness is the only site we study where it is unreasonable to assume skew surge and peak tide are independent~\citep{DArcy22}. The highest astronomical tide (HAT) observed is 10.72m, 2.92m, 6.10m and 6.26m for Heysham, Lowestoft, Newlyn and Sheerness, respectively. A linear mean sea level trend was removed from the data at each site therefore all of our results are presented relative to the mean sea level in the year 2017.~\cite{CFB2018} details this preprocessing stage; they remove estimated linear trends of 1.52, 2.27, 1.73 and 1.81mm per year from Heysham, Lowestoft, Newlyn and Sheerness, respectively. Of course, these trends incorporate land level changes as well as climate caused sea level changes, and also are based on different time periods as they correspond to the sample record at each site.

\subsection{Extreme Value Inference}\label{sec::EVA}
Within extreme value inference, it is natural to first consider modelling the maximum of a sequence $M_n=\max\{Z_1, \ldots, Z_n\}$. We first assume this sequence is independent and identically distributed (iid) with marginal distribution $F$ and upper end point $x^F$. If there exists sequences of constants $\{a_n>0\}$ and $\{b_n\}$ so that the rescaled block maximum $(M_n-b_n)/a_n$ has a nondegenerate limiting distribution as $n\rightarrow\infty$, then the distribution function $G$ of the maximum must be of the form \begin{equation}
    G(z)=\exp\bigg\{-\bigg[1+\xi\bigg(\frac{z-\mu}{\sigma}\bigg)\bigg]^{-1/\xi}_{+}\bigg\},\label{eqn::GEV} 
\end{equation} where $x_+=\max\{x,0\}$ whatever the distribution function $F$. This distributional model $G$ has three parameters $\mu\in\mathbb{R}$, $\sigma\in\mathbb{R}_{+}$ and $\xi\in\mathbb{R}$ representing the location, scale and shape, respectively~\citep{Coles2001}. This is known as the generalised extreme value distribution (GEV). For $\xi>0$ this corresponds to the Fr\'echet distribution, $\xi<0$ the Weibull and $\xi=0$ the Gumbel (although $\xi=0$ should be interpreted as the limit as $\xi\rightarrow 0$). This result, often referred to as the \textit{extremal types theorem}, gives an asymptotic justification to use the GEV as a model for block maxima, often taken to be annual maxima in environmental applications. However, in these settings, an iid assumption is usually unrealistic. A more commonly accepted assumption is stationarity, where the series can exhibit mutual dependence, but the statistical properties are homogeneous through time. If we now assume that $Z_1, \ldots, Z_n$ are from a stationary series that satisfies a long-range dependence condition, so that events far enough apart in time are near independent. Under these conditions this limiting distribution must be of the form $G^\theta(z)$ with $G(z)$ in expression~\eqref{eqn::GEV} and $\theta\in(0,1]$ the extremal index~\citep{Leadbetter1983}. 

If a process exhibits dependence, values above a high threshold $z$ form clusters, for example during a storm that spans multiple days we might observe several extreme skew surge values consecutively. We identify clusters as those separated by a run length $r$ defined as the number of consecutive observations below the high threshold $z$, i.e., \textit{non-extreme} values. Choosing this run length can be subjective, though~\cite{FerroSegers2003} propose an automated selection procedure based on the distribution of all times between consecutive exceedances of $z$. We can reasonably assume that observations in different clusters are independent, but this is not the case for observations in the same cluster. The extremal index $\theta$ provides information about clusters because it can be empirically estimated (known as the runs estimate) as the reciprocal of the mean cluster size~\citep{SmithWeissman1994}. These are both actually estimates of the subasymptotic extremal index \begin{equation}
    \theta(z,r)=\mathbb{P}(\max\{ Z_2,\ldots, Z_r\}<z| Z_1>z).\label{eqn::subas_exi}
\end{equation} Then the extremal index is defined as the limit of expression~\eqref{eqn::subas_exi} as $z\rightarrow z^F$ and $r\rightarrow\infty$ in a related fashion~\citep{LedfordTawn2003}.

An alternative, and more popular, approach to defining extreme values is as exceedances of a high threshold $u$. If the extremal types theorem holds, then for an arbitrary term $Z$ in the series $Z_1,\ldots,Z_n$,\begin{equation}
    \mathbb{P}(Z>b_n+a_n z\;|\;Z>a_n+b_n u)\rightarrow H_u(z) \quad \text{ where } \quad
    H_u(z)=\bigg[1+\xi\bigg(\frac{z-u}{\sigma_u}\bigg)\bigg]_+^{-1/\xi}\label{GPD_form}
\end{equation} for $z>u$ as $n\rightarrow\infty$, with $\{a_n>0\}$ and $\{b_n\}$ sequences of constants and $H_u$ is non-degenerate. This is known as the generalised Pareto distribution (GPD) and has two parameters $\sigma_u\in\mathbb{R}_{+}$ and $\xi\in\mathbb{R}$ representing the scale and shape, respectively~\citep{Coles2001}. Notice the scale parameter is threshold dependent since $\sigma_u=\sigma+\xi(u-\mu)$ for $\mu$ and $\sigma$ the GEV parameters; the shape parameter is the same as that for the GEV. Assuming $Z_1,\ldots,Z_n$ are iid, exceedances of a high threshold $u$ are also iid and have limiting GPD tail model \begin{equation}
    \mathbb{P}(Z>z)=\lambda_u\bigg[1+\xi\bigg(\frac{z-u}{\sigma_u}\bigg)\bigg]_+^{-1/\xi}
\end{equation} for $z>u$ where $\lambda_u=\mathbb{P}(Z>u)$. We can write the mean of excesses of the threshold $u$ as \begin{equation}
    \mathbb{E}(Z-u|Z>u)=\frac{\sigma_u}{1-\xi}.\label{eqn::mean_excess}
\end{equation}
However, if $Z_1,\ldots,Z_n$ are a dependent stationary series, a common approach is to identify clusters and decluster them (e.g., by considering cluster maxima only) to yield an approximately independent sequence so that the asymptotic justification for the GPD remains valid (\cite{SmithTawnColes1997},~\cite{FawcettWalshaw2007}). We subsequently drop the $u$ subscript on the scale $\sigma$ and rate $\lambda$ parameters.

\subsection{Existing Methodology}\label{sec::ex_methods}
The earliest methods directly modelled sea levels, but this ignores the known tidal component \citep{graff1978concerning, Coles1999, tawn1988sealevels}. \cite{DixonTawn1999} demonstrate that these approaches underestimate return levels. \cite{PughVassie1978} were the first to exploit the components of sea levels in their joint probabilities method (JPM) using surge and tide. \cite{Tawn1992} presents the revised joint probabilities method (RJPM) to address limitations associated with the JPM; mainly, they use an extreme value distribution to model the upper tail of surges to allow extrapolation beyond the range of observed values and, through a parameteric model, attempt to account for tide-surge dependence. The main pitfall with both of these approaches is that surge and tide have a complex joint distribution which is difficult to model effectively. \cite{Batstone2013} proposed the skew surge joint probabilities method (SSJPM) to avoid this complexity. This uses skew surge and peak tide as two components of sea levels, since they have a much weaker dependence and can be reasonably assumed to be independent at most sites on the UK National Tide Gauge Network~\citep{Williams2016}. \cite{baranes2020} build on this by accounting for interannual tidal variations and considering separate distributions for summer and winter skew surges; this is the quasi non-stationary skew surge joint probabilities method (qn-SSJPM).  

We build on the sea level model presented by~\cite{DArcy22} that uses skew surge and peak tide as two components of sea levels in a joint probabilities framework. This was first approach to capture the within year seasonality of each component and the dependence between them by adding covariates to the model parameters. It also accounts for skew surge temporal dependence which addresses previous issues of overestimation at short return periods. We describe their model for the annual maxima sea levels $M$. For a tidal cycle $i$, the peak sea level $Z_i$ can be written as the sum of the deterministic peak tide $X_i$ and stochastic skew surge $Y_i$. We first present their skew surge model, then describe how this is combined with the known peak tides to derive a sea level distribution. Lastly, we detail their model for the extremal index of skew surges used to derive the annual maxima distribution.

Since extreme sea levels can occur with various combinations of skew surge and peak tide, it is important to have a model for the entire skew surges distribution. To split the distribution into the body and tail, they use a monthly threshold $u_j$ for $j=1,\ldots,12$ to account for seasonality, with $u_j$ being a quantile, for a fixed percentile, of month $j$'s skew surge distribution. This choice ensures a similar number of exceedances each month. They use the 0.95 percentile, this is chosen based on monthly parameter stability plots~\citep{Coles2001}. Skew surges below these thresholds are modelled using the monthly empirical distribution $\tilde F_{j,x}$ to capture within year non-stationarity, that is also dependent on tides $x$ to account for skew-surge peak tide dependence. This empirical distribution is split into three associated peak tide bands: 
\begin{equation}
    \tilde F_{j,x}(y)=\begin{cases}
            \tilde F_{j}^{(1)}(y) \quad \text{if } x\leq x_{0.33}^{(j)} \\ 
            \\
            \tilde F_{j}^{(2)}(y) \quad \text{if } x_{0.33}^{(j)}<x\leq x_{0.67}^{(j)} \\
            \\
            \tilde F_{j}^{(3)}(y) \quad \text{if } x>x_{0.67}^{(j)},
\end{cases}\label{eqn::emp_ss_dist}
\end{equation} 
where $x_q^{(j)}$ denotes the $q$ quantile of the peak tide distribution for month $j$ and $\tilde F_{j}^{(l)}$ for $l=1,2,3$ is the empirical distribution of skew surges in month $j$ which are associated with the lowest ($l=1$), medium ($l=2$) and highest ($l=3$) band of peak tides. Since tide gauges on the UK National Tide Gauge Network usually have long observational periods, this can reliably model the main body of the data. For exceedances of the monthly threshold, they use a non-stationary GPD dependent on day in year $d=1,\ldots,365$, month $j$ and peak tide $x$. Therefore, the full skew surge model is given by \begin{align}
    F_Y^{(d,j,x)}(y)=\begin{cases}
            \tilde F_{j,x}(y) \quad &\text{if } y\leq u_j\\
            \\
            1-\lambda_{d,x}\big[1+\xi\big(\frac{y-u_j}{\sigma_{d,x}}\big)\big]_+^{-1/\xi} \quad &\text{if } y>u_j,
    \end{cases}\label{eqn::ss_model}
\end{align} where $\lambda_{d,x}$, $\sigma_{d,x}$ and $\xi$ are parametric functions to be estimated. Notice that the shape parameter $\xi$ does not vary with any covariate; this is kept fixed to avoid introducing additional uncertainty associated with estimating this parameter. The rate and scale parameters both depend on day and peak tide. They model the scale parameter using a harmonic for seasonal variations and a linear trend in terms of tide, \begin{equation}
    \sigma_{d,x}=\alpha_\sigma+\beta_\sigma\sin\bigg(\frac{2\pi}{f}(d-\phi_\sigma)\bigg)+\gamma_\sigma x,\label{eqn::ss_scale}
\end{equation} 
for $\alpha_\sigma>\beta_\sigma>0$ , $\phi_\sigma\in[0,365)$, $\gamma_\sigma\in\mathbb{R}$ parameters to be estimated and $f=365$ the periodicity.
The rate parameter is modelled similarly, using a generalised linear model with logit link function and a harmonic to capture seasonal variations. They also use a harmonic to capture how skew surge-peak tide dependence changes with time; \cite{DArcy22} show that this relationship varies throughout the year at Sheerness, with the strongest dependence occurring in May. This parameterisation is given by 
\begin{equation}
    g(\lambda_{d,x})=g(\lambda)+(d_j-\bar d_j)\beta_\lambda^{(d)}\sin\bigg(\frac{2\pi}{f}(d-\phi_\lambda^{(d)})\bigg)+\bigg(\frac{x-\bar x}{s_x}\bigg)\bigg[\alpha_\lambda^{(x)}+\beta_\lambda^{(x)}\sin\bigg(\frac{2\pi}{f}(d-\phi_\lambda^{(x)})\bigg)\bigg], \label{eqn::ss_rate}
\end{equation}
for $g(\cdot)$ the logit link function, $\lambda$ the constant exceedance probability in a month, $d_j\in[1,31]$ day in month (standardised by the monthly mean day $\bar d_j$), $\bar x$ is the mean and $s_x$ the standard deviation of peak tides, and $\alpha_\lambda^{(x)}\in\mathbb{R}$, $\beta_\lambda^{(d)},\beta_\lambda^{(x)}>0$, $\phi_\lambda^{(d)},\phi_\lambda^{(x)}\in[0,365)$ are parameters to be estimated. 

To derive a distribution for the sea levels, \cite{DArcy22} use a joint probabilities method and the fact that peak tides are deterministic. So that \begin{equation}
    \mathbb{P}(Z_i\leq z)=\mathbb{P}(X_i+Y_i\leq z)
=\mathbb{P}(Y_i\leq z-X_i)=F_{Y}(z-X_i), \mbox{ for }-\infty<z<\infty.
\end{equation} Let $T_j^{(k)}$ denote the number of tidal cycles in month $j$ and year $k$. They capture within and across year peak tide non-stationarity by using sequential monthly and yearly peak tide samples $X_{j_i}^{(k)}$, so that $j_i$ denotes the $i$th peak tide in month $j$, and $k=1,\ldots,K$ represents the year. Since peak tides are temporally dependent, the samples $\{X_{j_i}^{(k)}\}$ are from contiguous peak tides. Then the distribution of the annual maxima sea level $M$ is 
\begin{equation}
    \mathbb{P}(M\leq z)=\frac{1}{K}\sum_{k=1}^K\prod\limits_{j=1}^{12}\prod\limits_{i=1}^{T_j^{(k)}}F_{Y}^{(d,j,x)}(z- X^{(k)}_{j_i})^{\theta(z-X_{j_i}^{(k)},r)}
\label{eqn::monmax_dist}
\end{equation} where $F_{Y}^{(d,j,x)}$ is the skew surge model~\eqref{eqn::ss_model} and $\theta(z-X_{j_i}^{(k)},r)$ is a model for the extremal index, dependent on skew surge level $y=z-X_{j_i}^{(k)}$ and run length $r$, to capture temporal dependence of skew surges. This model is given by \begin{equation}
    \hat\theta(y,r)=\begin{cases}
        \tilde\theta(y,r) \quad\quad &\text{if }y\leq v \\
        \theta-[\theta-\tilde\theta(v,r)]\exp\big(-\frac{y-v}{\psi}\big) \quad\quad &\text{if }y>v,
    \end{cases}\label{eqn::exi_model}
\end{equation} where $v$ is a high threshold (they take the 0.99 percentile), $\psi>0$ and $\tilde\theta(v,r)\leq\theta\leq 1$ are parameters to be estimated and $\tilde\theta(y,r)$ is the empirical runs estimate. 


\subsection{Incorporating Interannual Variations to Skew Surge Distribution}\label{sec::trends}
We provide a framework to explore longer-term trends in extreme skew surges that may result from an increase in storm frequency and intensity. After removing the mean sea level trend, we follow the approach of~\cite{EastoeTawn2009} by adding yearly and global mean temperature anomaly (GMT) covariates into the scale and rate parameters to the GPD model for extreme skew surges presented by~\cite{DArcy22}. We do not consider adding covariates to the shape parameter or to the empirical distribution used for non-extreme skew surges. Another option would be to add covariates to the threshold choice, but it is difficult to account for the uncertainty in the threshold selection in extreme value inference~\citep{wadsworth2012, northrop2017}. Since we are interested in the temporal change in extreme events, it seems problematic to allow the threshold to also vary with time.

The model of \cite{DArcy22} already accounts for short term variations in the threshold exceedance rate and the GPD scale parameter. So, here we are focusing on the additional long term changes in these two features, knowing that estimates of these are not contaminated by short term features. Trends in the two features tell us about different aspects of the occurrence of  extreme skew surge events. An increase in the threshold exceedance rate tells us simply that more extreme events are occurring over time or with GMT increases. In contrast, increases in the scale parameter inform us that the nature of the extreme events are changing, in that their average size is getting larger. So it is of interest to explore both aspects.

First we consider how the threshold exceedance probability varies with year $k$ and GMT in year $k$ measured in $^\circ C$, denoted $m_k$. This tells us how the frequency of extreme skew surges is changing in response to climate change. We refer to the model for $\lambda_{d,x}$ introduced by~\cite{DArcy22} as $R0$, given by~\eqref{eqn::ss_rate}. We propose four model extensions of R0 to account for how the threshold exceedance rate also changes with $k$ (Models~$R1$ and~$R2$) and with $m_k$ (Models~$R3$ and~$R4$); with the odd numbered models having a single trend across the year and the even numbered models having a different linear trend per season,  with seasons $\{\mathcal{S}_s,
s=1,2,3,4\}$ denoting winter (December, January, February), spring (March, April, May), summer (June, July, August) and autumn (October, November December), respectively. These models are parametrised as follows,
\begin{align}
   & \text{Model~}R1\text{: } \quad g(\lambda_{d,x,\tilde k})=g(\lambda_{d,x})+\delta_{\lambda}^{(\tilde k)}\tilde k,  \label{eqn::mod_lambda_yr}\\
   & \text{Model~}R2\text{: } \quad g(\lambda_{d,x,\tilde k})=g(\lambda_{d,x})+ \sum\limits_{s=1}^{4}\delta_{\lambda,s}^{(\tilde k)}\tilde k \bm{1}_{\{d\in \mathcal{S}_s\}}, \label{eqn::mod_lambda_ssnl_yr}\\
   & \text{Model~}R3\text{: } \quad g(\lambda_{d,x,m_k})=g(\lambda_{d,x})+\delta_{\lambda}^{(m)}m_k, \label{eqn::mod_lambda_gmt}\\
   & \text{Model~}R4\text{: } \quad g(\lambda_{d,x,m_k})=g(\lambda_{d,x})+ \sum\limits_{s=1}^{4}\delta_{\lambda,s}^{(m)}m_k \bm{1}_{\{d\in \mathcal{S}_s\}},
   \label{eqn::mod_lambda_ssnl_gmt}
\end{align} where $g(\cdot)$ is the logit link function, which is selected to help our modelling of probabilities with linear models,
$\delta_{\lambda}^{(\tilde k)}, \delta_{\lambda,s}^{(\tilde k)}, \delta_{\lambda}^{(m)}, \delta_{\lambda,s}^{(m)}\in\mathbb{R}$ are parameters to be estimated ($s=1,2,3,4$), and $\tilde k\in\mathbb{R}$ denotes the standardised year defined as $\tilde k=(k-1968)/53$ where $k$ is the year of observation. Here 1968 is the midpoint of all years of observation and 53 is half of the range. We standardise in this way so that the results are comparable across sites with different observation lengths. For our study period, the covariates take values $\tilde k\in[-1,1]$ and $m_k\in(-0.56, 0.94)$; recall GMT is an anomaly centred at the temperature in the period 1961-1990, so it has been standardised somewhat already.

We consider these four models to investigate whether time or GMT is the best linear predictor of extreme skew surge non-stationarity over our observation period, and to explore if
the longer-term trends are non-stationary within a year, for example, if extreme skew surges are becoming more frequent in the winter but less so in summer. For Model $R1$, we are particularly interested in the change in 
threshold exceedance probability over the period 1920-2020 (100 years), this is given by $\Delta_\lambda^{(\tilde k)}=\lambda_{d,x,b}-\lambda_{d,x,a}$, for $a=-0.91$ (1920) and $b=1$ (2020). Similarly for Model $R3$, we define the change in exceedance probability with an increase in GMT of $1^\circ$C as $\Delta_\lambda^{(m)}=\lambda_{d,x,1}-\lambda_{d,x,0}=\lambda_{d,x,1}-\lambda_{d,x}$.

Next, we investigate how the GPD scale parameter changes with year and GMT to understand if the magnitude of extreme events is changing due to climate change. We extend the $\sigma_{d,x}$ parameterisation~\eqref{eqn::ss_scale} of~\cite{DArcy22} (call this Model~$S0$) and consider four models that capture changes with year, GMT and season as we did for 
the threshold exceedance rate,
\begin{align}
    & \text{Model~}S1\text{: } \quad \sigma_{d,x,k} = \sigma_{d,x} + \delta_{\sigma}^{(\tilde k)}\tilde k, \label{eqn::mod_sig_yr}\\
    & \text{Model~}S2\text{: } \quad \sigma_{d,x,k} = \sigma_{d,x} + \sum\limits_{s=1}^{4}\delta_{\sigma,s}^{(\tilde k)}\tilde k \bm{1}_{\{d\in \mathcal{S}_s\}}, \\
     & \text{Model~}S3\text{: } \quad \sigma_{d,x,m_k} = \sigma_{d,x} + \delta_{\sigma}^{(m)}m_k, \\
    & \text{Model~}S4\text{: } \quad \sigma_{d,x,m_k} = \sigma_{d,x} + \sum\limits_{s=1}^{4}\delta_{\sigma,s}^{(m)}m_k \bm{1}_{\{d\in \mathcal{S}_s\}}
    \label{eqn::mod_sig_ssnl_gmt},
\end{align} 
with parameters $\delta_{\sigma}^{(\tilde k)}, \delta_{\sigma,s}^{(\tilde k)}, \delta_{\sigma}^{(m)}, \delta_{\sigma,s}^{(m)}\in\mathbb{R}$ to be estimated and $\tilde k$, $m_k$, $\mathcal{S}_s$ as in \eqref{eqn::mod_lambda_yr}-\eqref{eqn::mod_lambda_ssnl_gmt}.

\subsection{Spatial Pooling}\label{sec::dependence}
\subsubsection{Improved Inference by Pooling}
So far we have described the modelling of extreme skew surges at a single site. However, this approach can be very inefficient, particularly for sites with short records or where the physical processes exhibit similarly over the sites, e.g., the same storm events effect all of the different sites. In such cases we would anticipate certain parameters of the extreme surge skew distribution to be similar, or even identical, in value across sites. By imposing this feature into the inference and carrying out joint inferences across sites, known as pooling, this can lead to large improvements in parameter estimation, by effectively sharing information about extreme events across sites, which in turn reduces estimation uncertainty resulting in narrower confidence intervals.

In the model of \cite{DArcy22} the benefits of pooling was illustrated for the shape parameter. This parameter is known to very difficult to estimate with much precision, with the variability in its estimator being the primary source of uncertainty in return level estimation.
This parameter has been recognised across a wide spectrum of problems as being very similar for a given process over large spatial regions, e.g., for rainfall~\citep{davison2012}, sea levels~\citep{dixontawnvassie1998} and air temperature~\citep{huser2016} with different values for the shape parameter for plains and mountains.~\cite{DArcy22} use information from \cite{CFB2018} that the shape parameter estimates, estimated separately from each site over the UK, followed a normal distribution with mean $0.0119$ and variance $0.0343$. They account for this in the likelihood inference, using this distribution as a prior penalty function. \cite{DArcy22} obtained shape parameter estimates which were more similar over sites with much reduced uncertainty, thus resulting in uncertainty reduction of high return level estimates. For example, for the 10,000 year return level at Sheerness, the 95\% confidence interval was reduced by 2.5m, corresponding to a factor of 6.

In our context the difficult parameters to estimate are those of the longer-term trends in  expressions~\eqref{eqn::mod_lambda_yr}-\eqref{eqn::mod_lambda_ssnl_gmt} for the threshold exceedance rate and~\eqref{eqn::mod_sig_yr}-\eqref{eqn::mod_sig_ssnl_gmt} for the GPD scale parameter. Here we also want to share spatial information through pooling. Given that we do not know if these trend parameters are identical over sites, and we only are illustrating the method for four sites, we undertake a formal likelihood testing method to assess the evidence to see if we can treat these trend parameters as constant over sites, without reducing the quality of the fit relative to the improved parsimony.

The pooled inference procedure involves a combined likelihood function
$L({\bm\theta})$ which combines the likelihood functions $L_i({\bm\theta}_i)$ from each of the $i=1, \ldots ,4$ sites through
\[
L({\bm\theta})=
\prod_{i=1}^{4} L_i({\bm\theta}_i),
\]
where ${\bm\theta}_i$ are the parameters for site $i$ and
${\bm\theta}=({\bm\theta}_1, \ldots ,{\bm\theta}_4)$. This likelihood enables hypothesis testing to be carried out, to assess the evidence for whether certain parameters are the same at all, or a subset set, of the sites, i.e., is the time trend gradient parameter the same at all sites. The joint likelihood function then enables the sharing of information about this common parameter across sites whilst allowing the other parameters to vary over sites. The choice of this joint likelihood function has potential restrictions; since it is a product over sites, this implicitly implies that extreme skew surges are being assumed to be independent across the sites. In cases where this assumption is unreasonable, the point estimates of the parameters will still be good (asymptotically consistent) but the variance of the estimates and the confidence intervals for the parameters will be underestimated, with the degree of underestimation dependent on the level of ignored true dependence there is between skew surges at the different sites. Therefore before exploiting the pooling strategy it is important to check that the independence assumption, for the extreme values of skew surge, is a reasonable approximation.

\subsubsection{Spatial Independence Diagnostics}
We discuss how to check for pairwise dependence between skew surges at different sites. Kendall's $\tau$ correlation coefficient can be used to check for dependence in all skew surge observations; this is a measure of rank correlation so is robust to outliers but it is a measure across all values of the variables. However, since our interest lies with the dependence of the extreme values, it is natural to also study the two main measures of extremal dependence $\chi$ and $\bar\chi$ \citep{Coles1999}, as described next.

Let $Y^{A}$ and $Y^{B}$ denote  skew surge random variables at two different sites $A$ and $B$, in the same tidal cycle with marginal distribution functions $F_A$ and $F_B$ respectively. The simplest measure of dependence is to see how the joint probability of $Y^{A}$ and $Y^{B}$ both being above their respective $(1-p)$th marginal quantiles, compares to $p$ (the value of this probability under perfect dependence of $Y^{A}$ and $Y^{B}$)
and relative to $p^2$ (the value under independence of $Y^{A}$ and $Y^{B}$). Under positive dependence we would expect that
\begin{equation}
    p^2<\mathbb{P}\{Y^{A}>F_A^{-1}(1-p), Y^{B}>F_B^{-1}(1-p)\}<p.\label{eqn:ordering}
\end{equation}
\cite{Coles1999} formalise this intuition to define the measure of extremal dependence as $p\rightarrow 0$, i.e., as we look above increasing quantiles. Specifically they take
\[
    \chi=\lim_{p\rightarrow 0}\mathbb{P}\{Y^{B}>F_B^{-1}(1-p)~\mid~ Y^{A}>F_A^{-1}(1-p)\}
\]
where $\chi\in[0,1]$. Increasing values of $\chi$ corresponding to stronger extremal dependence, and $\chi=1$ corresponds to perfect dependence between $Y^A$ and $Y^B$. Thus $\chi$ is the limiting  probability of one variable being extreme given that the other is equally extreme. If $\chi\in(0,1]$, we say $Y^A$ and $Y^B$ are asymptotically dependent, with there being a non-zero probability of $Y^B$ being large when $Y^A$ is large at extreme levels.  Though the class of extremal dependence where  $\chi>0$ is widely studied, this only corresponds to cases where the joint probability in~\eqref{eqn:ordering} is of  $O(p)$, i.e., decays as a multiple of $p$ as $p\rightarrow 0$. We find that $\chi=0$ in all other dependence cases as well as when $Y^A$ and $Y^B$ are actually independent, this class of variables is known as being asymptotically independent, and $\chi$ doesn't give us any information on the level of asymptotic independence. We need a more refined measure of extremal dependence than $\chi$ to enable us to separate between when there is some dependence of large values and independence of $Y^A$ and $Y^B$. \cite{Coles1999} also define 
\begin{equation}
    \bar\chi=\lim_{p\rightarrow 0} \frac{2\log\mathbb{P}\{Y^A>F_A^{-1}(1-p)\}}{\log\mathbb{P}\{Y^B>F_B^{-1}(1-p), Y^A>F_A^{-1}(1-p)\}}-1.
\end{equation} 
where $\bar\chi\in(-1,1]$. Here  $\bar\chi=1$ and $-1<\bar\chi<1$ correspond to asymptotic dependence and asymptotic independence, respectively. When $\bar\chi=0$ this shows there is no dependence in the tails of $(Y^A,Y^B)$ as it arises when $Y^A$ and $Y^B$ are independent, with  $0<\bar\chi<1$ 
and $-1<\bar\chi<0$ indicating positive and negative dependence in the joint tails of $Y^A$ and $Y^B$ respectively. 

To assess inter-site dependence in extreme skew surges we evaluate these dependence measures using empirical estimates of the associated probabilities using \texttt{texmex} R package~\citep{texmex}. Specifically, we use skew surge daily maxima for each pairwise combination of the four study sites, using data on the same day and with lags of $\pm 1$ day to account for time lags between the peak of  surge reaching each site, when events last multiple days. Here we have lags $t=1$ and $t=-1$ so that site $A$ is one day ahead or behind site $B$, respectively. Since the variables are not identically distributed, due to seasonality for example, this can effect the evaluation of $\chi$ and $\bar\chi$. We address this potential concern by also using the marginal distributional model of~\cite{DArcy22} $F_{Y}^{(d,j,k)}$, given by expression~\eqref{eqn::ss_model}, to account for this through a transform the variables to identical uniform margins and then re-evaluate these measures. 
These results are discussed in Section~\ref{sec::pooling_results}.


\section{Results}
\subsection{Introduction}\label{sec::results_intro}
We now present the results from applying the extreme skew surge models discussed in Sections~\ref{sec::trends} and~\ref{sec::dependence}, in Sections~\ref{sec::trends_results} and~\ref{sec::pooling_results}, respectively to data from our four study sites. Here we define extreme skew surges as being exceedances of the monthly 0.95 quantile, as in~\cite{DArcy22}. All models are fit in a likelihood framework, with 95\% confidence intervals provided for parameter estimates based on the hessian, i.e., using asymptotic normality of maximum likelihood estimators. The likelihood is constructed under the assumption that extreme skew surges are temporally independent for single site inference, but also that observations at different sites are independent for spatial pooling. These are not unreasonable assumptions for model selection, the former being found as a reasonable approximation in \cite{DArcy22} as the extremal index is near one for large levels and the validity of the latter being assessed  before we apply any spatial pooling. We compare models using AIC and BIC scores; these are commonly used measures of the quality of a statistical model for a particular data set relative to the parsimony of the model. The chosen best fitting model should minimise these scores. Recall that all estimates presented here are after the mean sea level trends have been removed. An estimated change here means that the change is the same as in the mean sea level, so negative trend estimates correspond to the extreme sea levels not rising as fast as the mean the level at the site.

\subsection{Single-site Analysis}\label{sec::trends_results}
We fit the models of Section~\ref{sec::trends} to the GPD rate and scale parameters for extreme skew surges individually at each site. We start with the threshold exceedance probability parameter, $\lambda$, fitting Models~$R0-4$; AIC/BIC scores and estimates of $\delta_\lambda^{(\tilde k)}$, $\delta_{\lambda,s}^{(\tilde k)}$, $\delta_{\lambda,s}^{(m)}$ and $\delta_\lambda^{(m)}$ are given in Table~\ref{tbl:rate_param_ests}. Since the parameter estimates are not intuitive, we consider the change in exceedance probability with the particular covariate of interest for the annual trends in Models~$R1$ and~$R3$. 

\begin{table}
    \centering
    \resizebox{\textwidth}{!}{
    \begin{tabular}{lcccc}
        \hline
         & Heysham & Lowestoft & Newlyn & Sheerness\\ \hline
         \multicolumn{5}{l}{Model $R0$} \\ \hline
         AIC & 12234.21 & 15312.08 & 24498.77 & 9286.58 \\
         BIC & \textbf{12275.89} & \textbf{15354.88} & 24543.93 & \textbf{9326.94} \\ \hline
         \multicolumn{5}{l}{Model $R1$} \\ \hline
         $\delta_\lambda^{(\tilde k)}$ & 0.091 (-0.008, 0.191) & -0.061 (-0.150, 0.028) & 0.215 (0.154,0.276) & -0.114 (-0.013, -0.010)\\
         AIC & 12232.89 & 15312.26 & 24453.12 & 9283.96 \\
         BIC & 12282.91 & 15363.61 & 24507.31 & 9332.40 \\ \hline
         \multicolumn{5}{l}{Model $R2$} \\ \hline
         $\delta_{\lambda,1}^{(\tilde k)}$ & 0.161 (-0.033, 0.335) & 0.063 (-0.106, 0.232) & 0.114 (-0.011, 0.238)& -0.032 (-0.228, 0.164) \\
         $\delta_{\lambda,2}^{(\tilde k)}$ & 0.034 (-0.161, 0.230) & -0.141 (-0.322, 0.040) & 0.197 (0.077, 0.316)& -0.250 (-0.468, -0.032) \\
         $\delta_{\lambda,3}^{(\tilde k)}$ & 0.207 (0.013, 0.400) & -0.094 (-0.266, 0.078) & 0.209 (0.089, 0.328) & -0.189 (-0.405, 0.026) \\
         $\delta_{\lambda,4}^{(\tilde k)}$ & -0.047 (-0.261, 0.167) & -0.081 (-0.264, 0.102) & 0.338 (0.217, 0.460) & -0.021 (-0.221, 0.178) \\
         AIC & 12235.30 & 15315.29 & 24452.52 & 9286.48 \\
         BIC & 12310.32 & 15392.33 & 24533.80 & 9359.14 \\ \hline
         \multicolumn{5}{l}{Model $R3$} \\ \hline
         $\delta_\lambda^{(m)}$ & 0.204 (0.074, 0.334) & -0.012 (-0.12, 0.427) & 0.336 (0.245, 0.427) & -0.164 (-0.304, -0.024)\\
         AIC & \textbf{12227.14} & \textbf{15314.04} & 24451.34 & \textbf{9283.20} \\
         BIC & 12277.16 & 15365.39 & \textbf{24505.53} & 9331.64 \\ \hline
        \multicolumn{5}{l}{Model $R4$} \\ \hline
         $\delta_{\lambda,1}^{(m)}$ & 0.256 (-0.002, 0.514) & 0.103 (-0.107, 0.312) & 0.135 (-0.058, 0.329) & -0.079 (-0.340, 0.181)  \\
         $\delta_{\lambda,2}^{(m)}$ & 0.111 (-0.143, 0.365) & -0.067 (-0.282, 0.149) & 0.322 (0.144, 0.501) & -0.374 (-0.672, -0.076) \\
         $\delta_{\lambda,3}^{(m)}$ & 0.416 (0.167, 0.665) & -0.048 (-0.259, 0.163) & 0.393 (0.212, 0.574) & -0.273 (-0.568, 0.022) \\
         $\delta_{\lambda,4}^{(m)}$ & 0.010 (-0.274, 0.293) & -0.040 (-0.262, 0.182) & 0.478 (0.300, 0.655) & -0.008 (-0.253, 0.269) \\
         AIC & 12227.92 & 15318.49 & \textbf{24450.27} & 9284.69 \\
         BIC & 12302.94 & 15395.53 & 24531.56 & 9357.34 \\ \hline
    \end{tabular}
    }
    \caption{Parameter estimates for the Models $R1-R4$ with AIC and BIC scores for each model fit at each site (including Model~$R0$). The minimum AIC/BIC scores are highlighted in bold for each site.}
    \label{tbl:rate_param_ests}
\end{table}

We find that Model~$R3$ minimises AIC at Heysham, Lowestoft and Sheerness, whilst at Newlyn Model~$R4$ is preferable. The BIC is minimised by Model~$R3$ at Newlyn, but elsewhere Model~$R0$ is favourable. This suggests that if any longer-term trends are included in the model to capture changes in the rate of extreme events (relative to mean sea level), GMT should be used as a covariate as opposed to the year.

\begin{figure}
    \centering
    \includegraphics[width=0.45\textwidth]{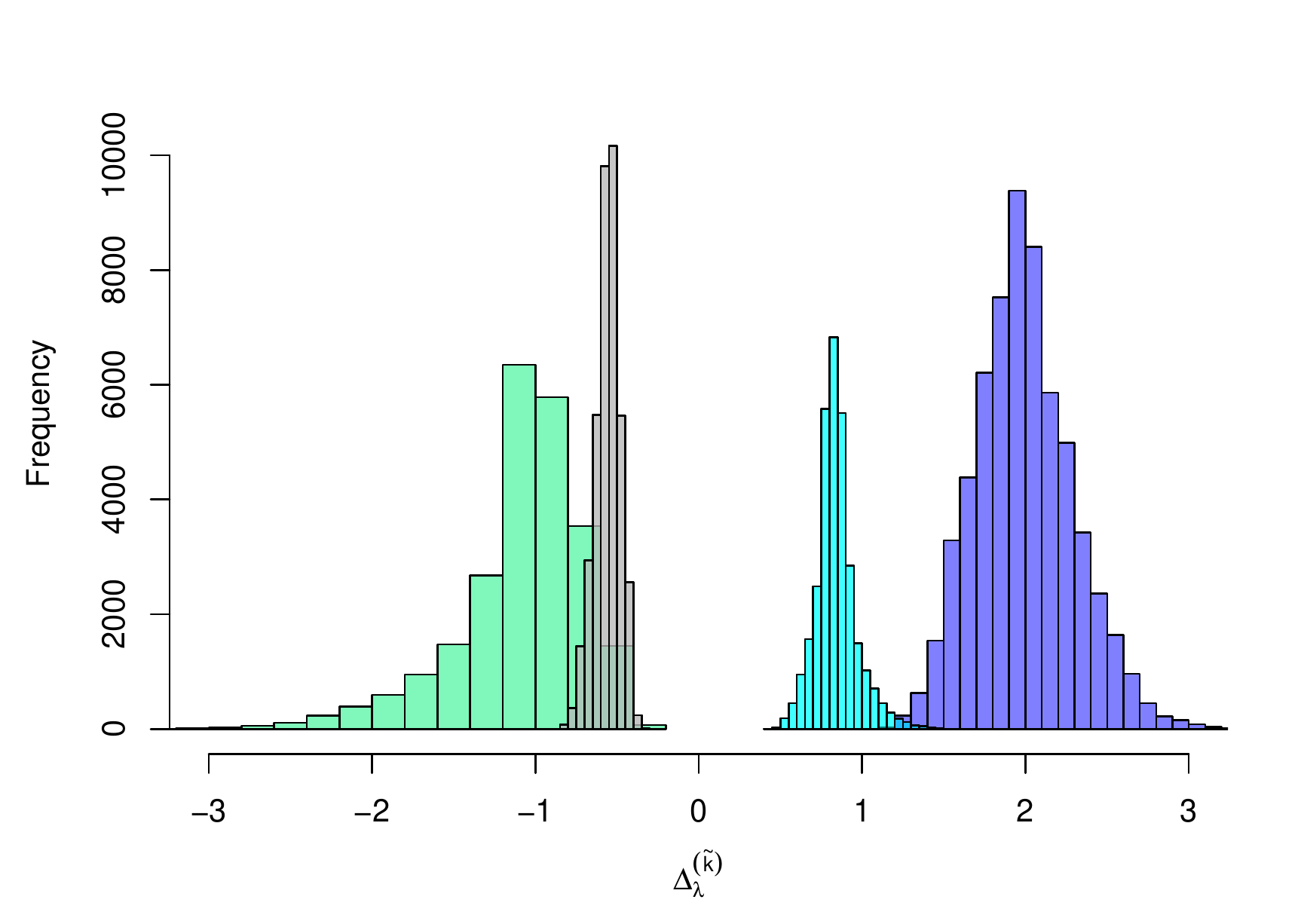}\includegraphics[width=0.45\textwidth]{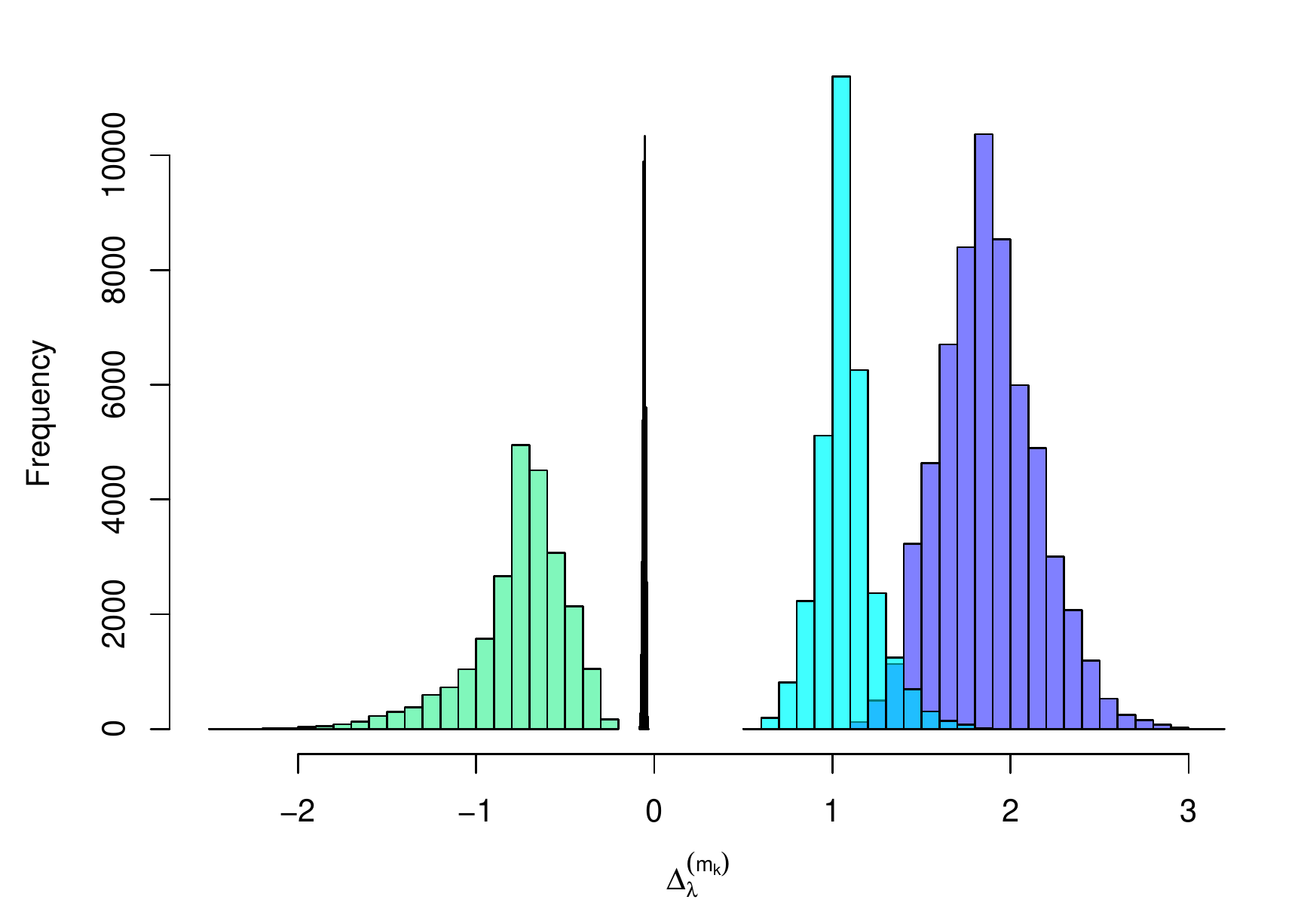}
    \caption{Histograms of $\Delta_{\lambda}^{(\tilde k)}$ over 100 years (left) and  $\Delta_\lambda^{(m)}$ with a $1^\circ$C increase in GMT, as percentages, for all day $d$ and peak tide $x$ combinations at Heysham (cyan), Lowestoft (grey), Newlyn (blue) and Sheerness (green).}
    \label{fig::hist_Delta}
\end{figure}

We look at Models~$R1$ and~$R3$ in more detail, these have a fixed trend parameter within the year with respect to year and GMT, respectively. At Newlyn we find a significant increasing trend for both models, since the confidence intervals do not contain zero. We also find positive trends at Heysham, but only the GMT trend in Model~$R3$ is significant. Neither trends are significant at Lowestoft, but we find a significant decreasing trend for both models at Sheerness. Figure~\ref{fig::hist_Delta} shows histograms of the estimates of $\Delta_\lambda^{(\tilde k)}$ and $\Delta_\lambda^{(m)}$ (defined in Section~\ref{sec::trends}), based on all combinations of day $d$ and peak tide $x$, so these do not account for uncertainty in $\delta^{(\tilde k)}_\lambda$ or $\delta_\lambda^{(m)}$ estimates but are simply a reflection that the rate of threshold exceedance varies over the short term. For Model~$R1$, we find an increase in $\lambda_{d,x,\tilde k}$ over 100 years at Newlyn, with $\max\Delta^{(\tilde k)}_{\lambda}=3\%$, so that the exceedance probability almost doubles from 3.5\% to 6.5\% in 1920-2020. However, we observe decreases in exceedance probability at Sheerness. For Model~$R3$, we also find an increase in exceedance probability with a $1^\circ$C increase in GMT at Newlyn, where $\max\Delta_{\lambda}^{(m)}=3\%$, but a negative trend at Sheerness. If the trends were statistically significant at Heysham and Lowestoft, the exceedance probability would be increasing and decreasing with both trend parameters, respectively.

\begin{figure}
    \centering
    \includegraphics[width=0.4\textwidth]{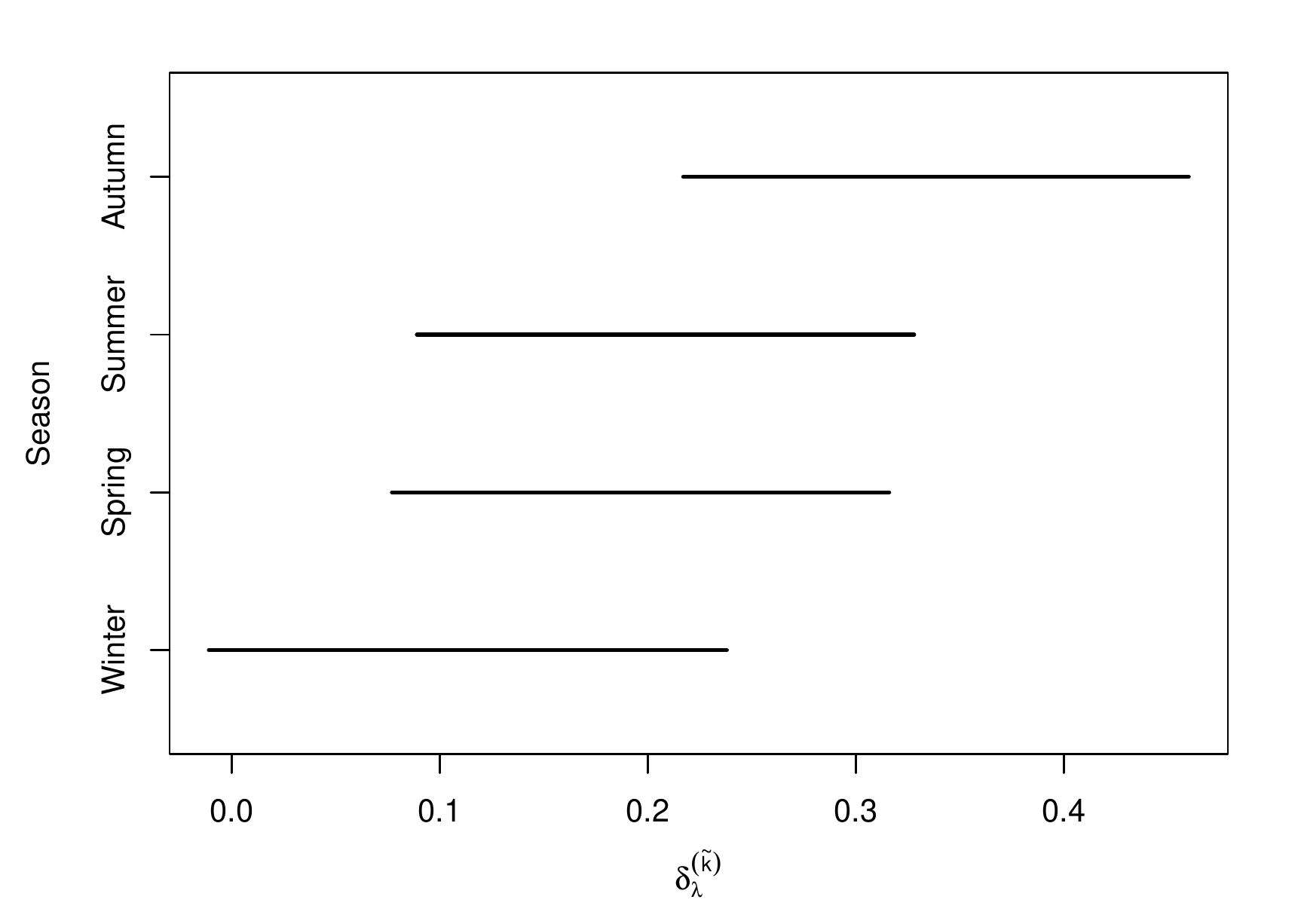}\includegraphics[width=0.4\textwidth]{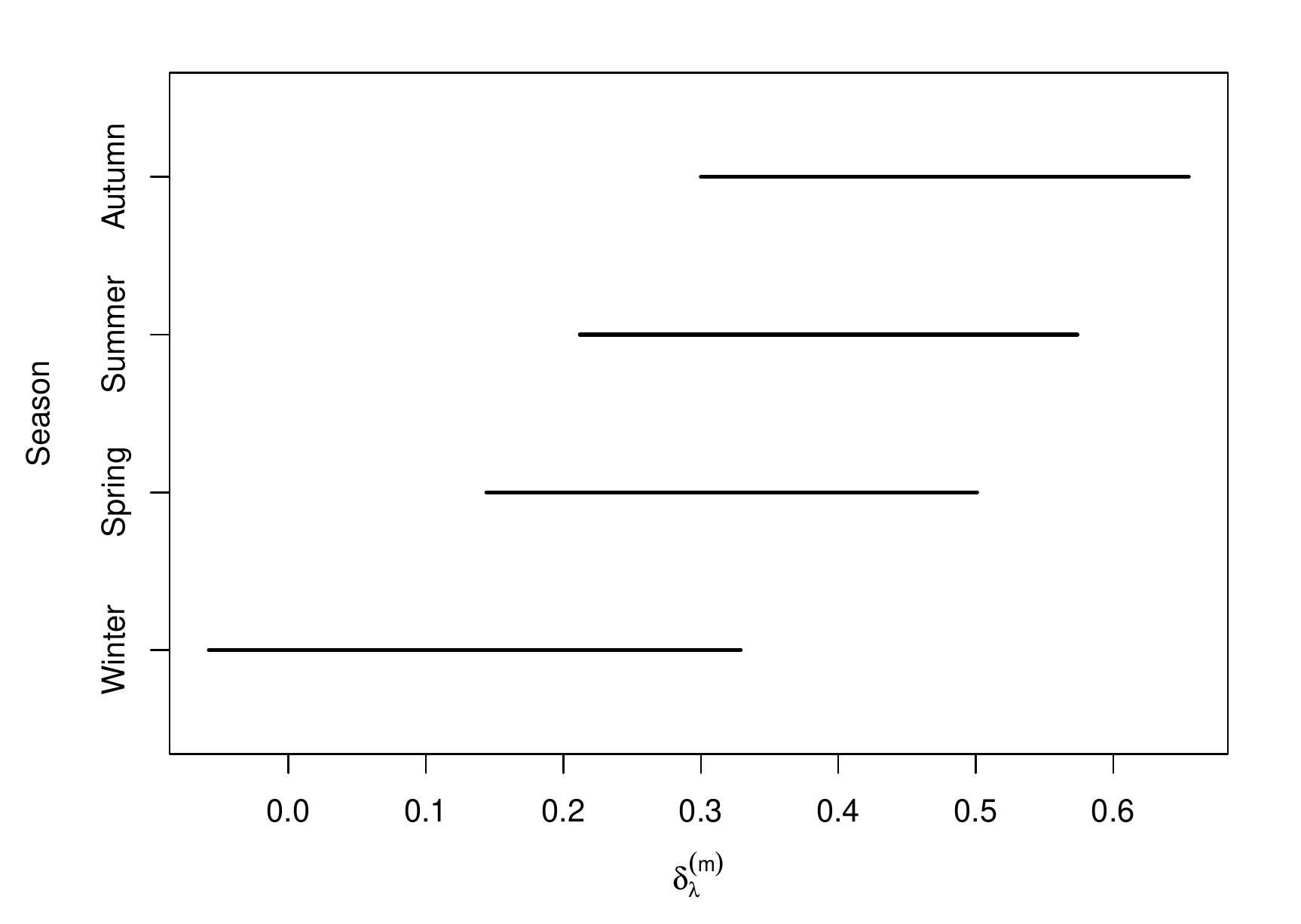}
    \caption{Confidence intervals for parameter estimates $\hat\delta_{\lambda,s}^{(\tilde k)}$ (left) and $\hat\delta_{\lambda,s}^{(m)}$ (right) at Newlyn for $s=1,2,3,4$ denoting winter, spring, summer and autumn, respectively.}
    \label{fig::Newlyn_R2_R4_CIs}
\end{figure}

Now we look at Models~$R2$ and~$R4$ with season-specific trend parameters for year and GMT, respectively. The trends at Newlyn are significant in both models, except for winter, whilst at Heysham only the increasing trends in summer are significant. None of the seasonal trends are significant at Lowestoft but we find a significant decreasing trend for spring in both models at Sheerness. As for Models~$R1$ and~$R3$, we obtain a variety of results across sites; Newlyn has an increasing exceedance probability with year and GMT in all seasons, whilst Sheerness has a decreasing trend parameter for all seasons, with the greatest decline in spring. However, for Heysham and Lowestoft, we have a mixture of positive and negative parameters throughout the year. The confidence intervals for the four parameter estimates in Models~$R2$ and~$R4$ at Heysham, Lowestoft and Sheerness overlap, suggesting that there isn't significant within-year variation of the longer-term trend parameters so that the simpler Models~$R1$ and~$R3$ are sufficient. Whilst at Newlyn, this overlap is small (see Figure~\ref{fig::Newlyn_R2_R4_CIs}). Here, we find the greatest trend in autumn, which is not concerning for extreme sea level estimation since the most extreme sea levels occur in winter~\citep{DArcy22}, but using Models~$R1$ and~$R3$ with common trend parameters across the year could result in overestimation of the trends in winter, hence influencing sea level return level estimation.

\begin{figure}
    \centering
    \includegraphics[width=0.4\textwidth]{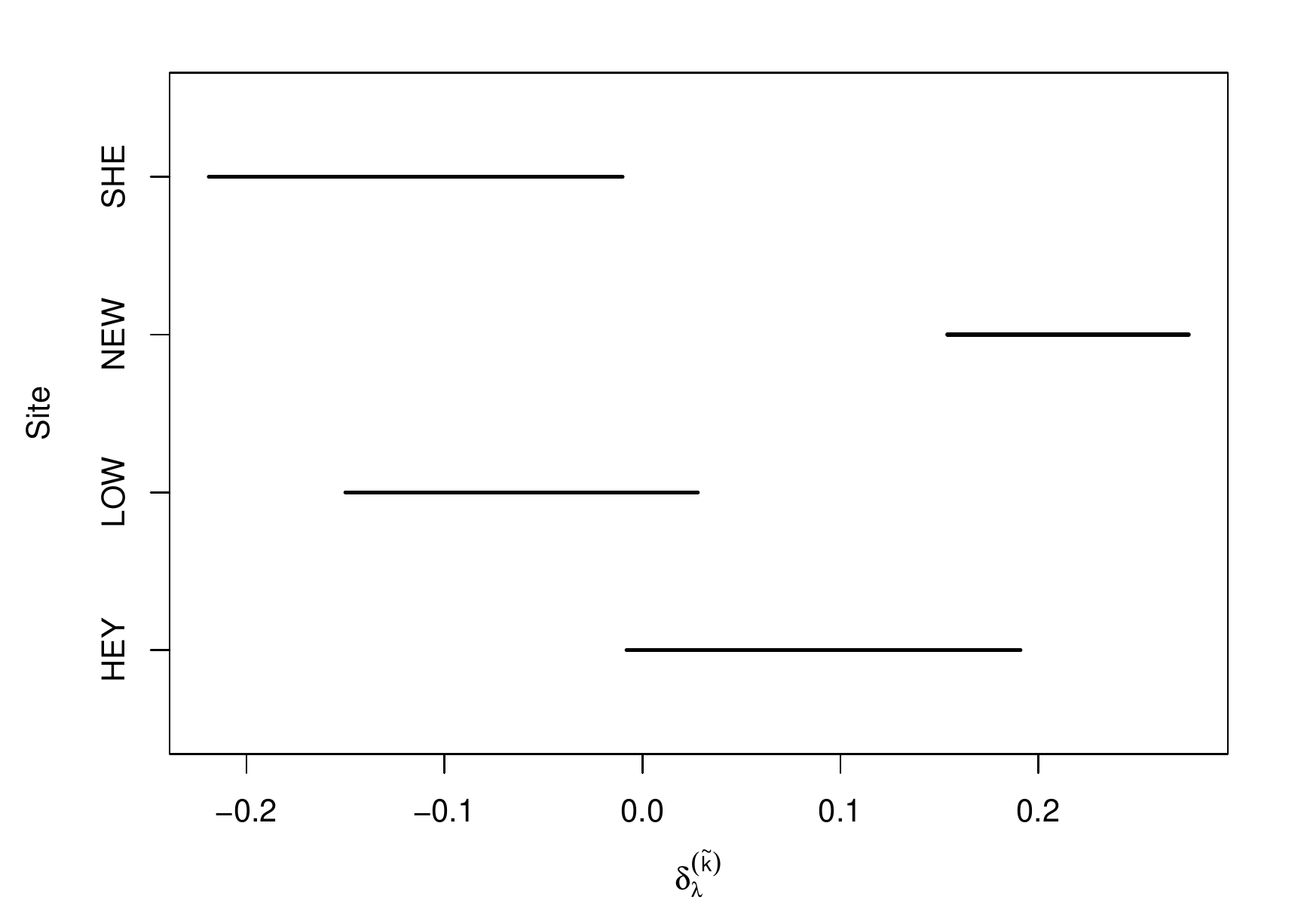}\includegraphics[width=0.4\textwidth]{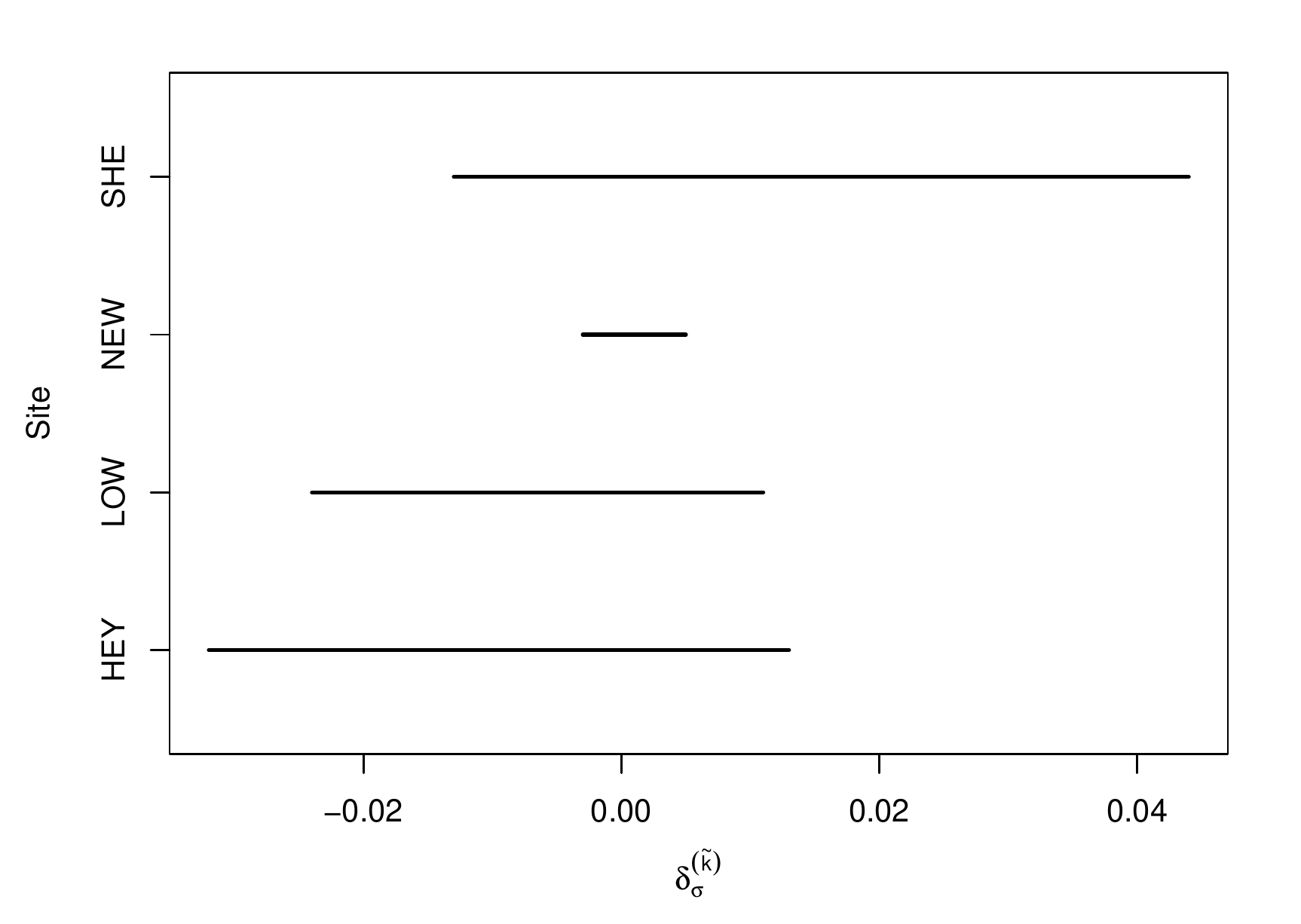}
    \caption{Confidence intervals for parameter estimates $\hat\delta_{\lambda}^{(\tilde k)}$ (left) and $\hat\delta_{\sigma}^{(\tilde k)}$ (right) for all sites.}
    \label{fig::Allsites_R1_S1}
\end{figure}

Next, we consider models for the scale parameter at each site individually (Models~$S0-4$, introduced in Section~\ref{sec::trends}). Table~\ref{tbl:scale_param_ests} shows the parameter estimates for each model, along with AIC and BIC scores. Models~$S1$ and~$S3$ have a single parameter denoting a common long-term trend across the year, but neither of these are an improvement on Model~$S0$ (without a long-term trend) at any site. All of the 95\% confidence intervals for $\delta^{(\tilde k)}_\lambda$ or $\delta_\lambda^{(m)}$ estimates contain zero, suggesting these trends are not significant. If we ignore this uncertainty, the point estimates suggest small changes in the scale parameter; at Heysham and Lowestoft our results show a decrease with both year and GMT, suggesting that the magnitude of extreme skew surge events are getting smaller with anthropogenic climate change effects. In the 100 year period 1920-2020 at Newlyn, the point estimate $\delta_\sigma^{(\tilde k)}$ corresponds to an increase in mean excesses (see expression~\eqref{eqn::mean_excess}) of 2mm (relative to a mean of 94mm in 1920), whilst at Sheerness in the years of observation 1980-2016 this corresponds to an increase of 10mm relative to a mean of 125mm in 1980. Notice there is overlap in the parameter estimates for $\delta_\sigma^{(\tilde k)}$ and $\delta_\sigma^{(m)}$ across sites; in Section~\ref{sec::pooling_results} we fit similar model with these trend parameters common across sites (see Figure~\ref{fig::Allsites_R1_S1}).

\begin{table}
    \centering
    \resizebox{\textwidth}{!}{
    \begin{tabular}{lcccc}
        \hline
         & Heysham & Lowestoft & Newlyn & Sheerness\\ \hline
         \multicolumn{5}{l}{Model $S0$} \\ \hline
         AIC & -3091.53 & -3672.07 & \textbf{-10152.63} &  \textbf{-2974.317} \\
         BIC & \textbf{-3064.77} & \textbf{-3644.20} & \textbf{-10122.42} &  \textbf{-2948.854} \\ \hline
         \multicolumn{5}{l}{Model $S1$} \\ \hline
         $\delta_\sigma^{(\tilde k)}$ & -0.009 (-0.032, 0.013) & -0.006 (-0.024, 0.011) & 0.001 (-0.003, 0.005) & 0.016 (-0.013,  0.044)\\
         AIC & -3088.558 & -3670.55 & -10150.80 & -2973.05  \\
         BIC & -3056.448 & -3637.11 & -10114.54 & -2942.50 \\ \hline
         \multicolumn{5}{l}{Model $S2$} \\ \hline
         $\delta_1^{(\tilde k)}$ & 0.022 (-0.034, 0.078) & -0.041 (-0.090, 0.007) & 0.004 (-0.009, 0.016) & 0.023 (-0.032, 0.078) \\
         $\delta_2^{(\tilde k)}$ & 0.022 (-0.014, 0.059) & -0.030 (-0.055, -0.006) & 0.006 (-0.002, 0.014)  & -0.001 (-0.036, 0.035) \\
         $\delta_3^{(\tilde k)}$ & -0.025 (-0.051, 0.001) & 0.012 (-0.010, 0.034) & -0.003 (-0.009, 0.003) & 0.023 (-0.010, 0.055) \\
         $\delta_4^{(\tilde k)}$ & -0.035 (-0.079, 0.008) & -0.015 (-0.053, 0.023) & 0.001 (-0.008, 0.011) & 0.008 (-0.039, 0.054) \\
         AIC & \textbf{-3095.28} & -3674.12 & -10146.27 & -2971.19  \\
         BIC & -3047.12 & -3623.96 & -10091.89  &  -2925.36 \\ \hline
         \multicolumn{5}{l}{Model $S3$} \\ \hline
         $\delta_\sigma^{(m)}$ & -0.011 (-0.033, 0.011) & -0.008 (-0.026, 0.009) & -0.001 (-0.008, 0.006) & 0.006 (-0.020, 0.032)\\
         AIC & -3088.42 & -3670.90 & -10149.07 & -2972.43 \\
         BIC & -3056.32 & -3637.46 & -10112.82 & -2941.87 \\ \hline 
        \multicolumn{5}{l}{Model $S4$} \\ \hline
         $\delta_1^{(m)}$ & 0.036 (-0.027, 0.099) & -0.050 (-0.105, 0.004) & -0.0003 (-0.021, 0.020) & 0.025 (-0.042, 0.091) \\
         $\delta_2^{(m)}$ & 0.029 (-0.012, 0.070) & -0.037 (-0.061, -0.012) & 0.005 (-0.008, 0.018) & -0.015 (-0.054, 0.023)  \\
         $\delta_3^{(m)}$ & -0.027 (-0.054, -0.00009) & 0.017 (-0.006, 0.039) & -0.003 (-0.013, 0.006)  & 0.013 (-0.018, 0.045) \\
         $\delta_4^{(m)}$ & -0.030 (-0.081, 0.021) & -0.024 (-0.066, 0.017) & -0.005 (-0.021, 0.010) & -0.006 (-0.053, 0.040) \\
         AIC & -3093.73 & \textbf{-3677.95} & -10145.39 & -2970.79  \\
         BIC & -3045.56 & -3627.79 & -10091.01 & -2924.96  \\ \hline
    \end{tabular}
    }
    \caption{Parameter estimates for the Models $S1-4$ with AIC and BIC scores for each model fit at each site (including Model~$S0$). The minimum AIC/BIC scores are highlighted in bold for each site.}
    \label{tbl:scale_param_ests}
\end{table}

Models~$S2$ and~$S4$ have four additional parameters relative to Model~$S0$, these denote a separate trend for each season with respect to year and GMT. AIC and BIC are still minimised by Model $S0$, except AIC scores for Heysham and Lowestoft, which favour Models~$S2$ and $S4$, respectively. However, the four confidence intervals overlap at each site, suggesting a fixed trend within a year is sufficient. At Heysham, the overlap across all seasons is small but there is considerable overlap between winter and spring, with positive trend parameters, and likewise for summer and autumn with negative trend parameters for both models. If these trends were statistically significant it would suggest that the magnitude of extreme skew surges is increasing with increases in GMT in December-April, but decreasing for the rest of the year. Given the timing of the most extreme events, this could be important if statistically significant.

\subsection{Spatial Pooling}\label{sec::pooling_results}
We present the results from pooling information across sites, for the long-term trend parameters, when refitting the models of Section~\ref{sec::trends}. Before pooling information, we use the dependence measures discussed in Section~\ref{sec::dependence} to check if it is reasonable to assume each pair of sites are independent in their extreme skew surge values. We estimate the dependence measures for the observed skew surges and a standardised transformation of them to remove sources of within-year non-stationarity via mapping to uniform margins through the distribution function~\eqref{eqn::ss_model}. The results are shown in Table~\ref{tbl::sitedep_ts}. For most combinations of sites at lags $t=-1, 0, 1$ the dependence is weak, with Kendall's $\tau$ of approximately $0.3$ for most pairs, except for Newlyn with the east coast site where Kendall's $\tau$ is near 0, whilst for Lowestoft and Sheerness this value is approximately 0.5. The effect of de-seasonalising the data (by transforming to uniform margins) has typically decreased dependence by at most $5\%$. Examining the extremal dependence measure $\chi$, we find that the de-seasonalised estimate is down from 0.3 to approximately 0.15 for most pairs, though unchanged for Lowestoft and Sheerness. With the exception of Lowestoft and Sheerness, it is not unreasonable to make an independence in extremes approximation for the data. Finally, looking at $\bar\chi$ we find $\bar\chi<1$ for all pairs, giving evidence of asymptotic independence with weak dependence in the observed tails of the variables. The strongest dependence is found between Lowestoft and Sheerness at lag $t=0$. This is not surprising since these sites are close in proximity, with extreme skew surges progressing south down the east coast through Lowestoft onto Sheerness. Therefore they are highly likely to be affected by the same storms. Despite this pair of sites giving clear evidence of dependence, we continue under the belief that it is reasonable to assume skew surge daily maxima at all pairs of sites are sufficiently close to being independent for the purposes of spatial pooling.

\begin{table}
    \resizebox{\textwidth}{!}{
    \centering
    \begin{tabular}{c||c|c|c||c|c|c||c|c|c||c|c|c||c|c|c||c|c|c}
        &  \multicolumn{3}{c||}{HEY-LOW} & \multicolumn{3}{c||}{HEY-NEW} & \multicolumn{3}{c||}{HEY-SHE} & \multicolumn{3}{c||}{LOW-NEW} & \multicolumn{3}{c||}{LOW-SHE} & \multicolumn{3}{c}{NEW-SHE}\\ \hline
        lag & -1 & 0 & 1 & -1 & 0 & 1 & -1 & 0 & 1 & -1 & 0 & 1 & -1 & 0 & 1 & -1 & 0 & 1 \\ \hline
        \multicolumn{19}{l}{Observations} \\ \hline
        $\tau$  & 0.133 & 0.160 & \textbf{0.309}  & 0.287 & \textbf{0.322} & 0.259  & 0.153 & 0.149 & \textbf{0.298}& \textbf{0.089} & 0.040  & 0.034 & 0.155 & \textbf{0.510} & 0.238 & 0.137 & 0.168 & \textbf{0.196} \\
        $\chi$ & 0.095 & 0.129  & \textbf{0.270} & 0.127 & \textbf{0.145} & 0.076& 0.092 & 0.111  & \textbf{0.259}& \textbf{0.017} & 0  & 0 & 0.145 & \textbf{0.509} & 0.200 & 0.054 & 0.077 & \textbf{0.121}\\
        $\bar\chi$ & 0.200 & 0.251 & \textbf{0.424} & 0.249 & \textbf{0.276} & 0.160  & 0.195 & 0.224 & \textbf{0.412} & \textbf{0.040} & -0.018  & -0.055 & 0.274 & \textbf{0.645} & 0.344& 0.120 & 0.158  & \textbf{0.237}\\ \hline
        \multicolumn{19}{l}{Transform to Uniform(0,1)} \\ \hline
        $\tau$ & 0.103 & 0.130 & \textbf{0.289} & 0.285 & \textbf{0.318} & 0.244 & 0.108 & 0.102 & \textbf{0.262}& \textbf{0.086} & 0.036  & 0.028 & 0.143 & \textbf{0.523} & 0.228 & 0.139 & 0.173 & \textbf{0.200} \\
        $\chi$ & 0.026 & 0.040 & \textbf{0.180} & 0.103 & \textbf{0.122} & 0.053 & 0.056 & 0.036 & \textbf{0.173} & 0 & 0 & 0 & 0.095 & \textbf{0.494} & 0.174 & 0.003 & 0.016 & \textbf{0.050} \\
        $\bar\chi$ & 0.069 & 0.100 & \textbf{0.321} & 0.215 & \textbf{0.236} & 0.114 & 0.123 & 0.084 & \textbf{0.313} & \textbf{-0.012} & -0.107 & -0.134  & 0.198 & \textbf{0.634} & 0.316 & 0.003 & 0.035 & \textbf{0.114}
    \end{tabular}
    }
    \caption{Kendall's $\tau$, $\chi$ and $\bar\chi$ measures of dependence for daily maximum skew surge observations at pairs of sites. We show the dependence over lags -1 (LHS site is 1 day behind RHS), 0 and 1 (LHS site is 1 day ahead of RHS); in bold we show the largest dependence over these lags. $\chi$ and $\bar\chi$ are measures of extremal dependence for exceedances of the 0.95 quantile.}
    \label{tbl::sitedep_ts}
\end{table}


Firstly, we focus on pooling information across sites regarding the long-term trend parameters with respect to year $k$ and GMT $m_k$. Figure~\ref{fig::Allsites_R1_S1} shows there is considerable overlap in the confidence intervals for $\hat\delta_{\lambda}^{(\tilde k)}$ at Lowestoft and Sheerness; similarly there is some overlap for Heysham and Newlyn. Although pooling information across randomly selected subsets of sites should be discouraged, here we note that the pairs of sites with similarities are on different coastlines, so we explore pooling over sites on the east coast (Lowestoft and Sheerness) and separately on the west coast (Heysham and Newlyn). Here, we consider refitting Models~$R1$ and~$R3$ (i.e., a fixed trend parameter within a year) with common trend parameters $\delta_{\lambda}^{(\tilde k)}$ and $\delta_{\lambda}^{(m)}$ between the pairs of sites. We obtain negative trends parameters $\hat\delta_{\lambda}^{(\tilde k)}=-0.084\;(-0.151, -0.016)$ and $\hat\delta_{\lambda}^{(m)}=-0.070\;(-0.156, 0.015)$ for Sheerness and Lowestoft, whilst at Newlyn and Heysham we obtain statistically significant positive trend parameters $\hat\delta_{\lambda}^{(\tilde k)}=0.180\;(0.128, 0.231)$ and $\hat\delta_{\lambda}^{(m)}=0.285\;(0.208, 0.359)$. Both of these models are an improvement on the previous results, where a separate long-term trend parameter is used for each site; the model with a yearly trend parameter reduces AIC by 48 and the BIC by 0.5, whilst the model with a GMT parameter reduces AIC and BIC by 54 and 6, respectively. This highlights the importance of sharing information spatially.


There is also information to be gained from sharing spatial information about long-term trends in the scale parameter since there is considerable overlap in the confidence intervals for $\hat\delta_{\sigma}^{(\tilde k)}$ (see Figure~\ref{fig::Allsites_R1_S1}) and $\hat\delta_{\sigma}^{(m)}$ (see Table~\ref{tbl:scale_param_ests}). We refit the models of Section~\ref{sec::trends} for the scale parameter with common longer-term trend parameters across sites, but neither parameter estimates are significant. We find that $\hat\delta_\sigma^{(\tilde k)}=4.8\times 10^{-4}$, corresponding to an increase in scale parameter of 0.48mm over 1915-2020. For GMT $\hat\delta_\sigma^{(\tilde k)}=-0.0024$, a 24mm decrease in scale parameter with a $1^\circ$C increase in temperature. Neither of these models improve the fit relative to having no long-term trends (in addition to those in mean sea level), although the AIC scores are close. We also fit a model similar to that of Models~$S2$ and~$S4$ so there is a common seasonal trend across sites, with respect to year and GMT but find that neither of these improve model fit. This agrees with our single-site results of Section~\ref{sec::trends_results} where we found no evidence of changes in the magnitude of extreme skew surge events with respect to year or temperature. 

\section{Discussion}\label{sec::discussion}
We have presented a framework to investigate the effects of anthropogenic climate change on extreme skew surges as any increases in the magnitude or frequency in these events can have catastrophic consequences if not included in extreme sea level estimation for coastal flood defence design. These trends can be different to those observed in the main body of the data, such as mean sea level rise. We use year and GMT as covariates in our statistical model for extreme event occurrence, building on a model developed by~\cite{DArcy22} that accounts for seasonality and skew surge-peak tide dependence. Recall that our results are relative to the mean sea level trend in 2017 so this would need to be added onto any sea level return level estimates when used in practice. We show that there is evidence of an increase in the probability of an extreme skew surge event with GMT increases at Heysham and Newlyn, but evidence of a decrease in the likelihood of these events at Lowestoft and Sheerness. We do not find any significant changes in the magnitude of extreme skews surges, i.e., in the scale parameter, and hence in the mean of the skew surge excesses of the threshold. We also looked at any seasonal variations in these longer-term trends and found evidence of a greater change in exceedance probability with year and GMT in autumn compared to the other seasons, at Newlyn. The ideas presented in this paper could be applied to more locations, but also to other environmental variables to investigate trends in extreme values.


We demonstrate the advantages of pooling information across sites, although this is only primarily illustrative since we consider just four sites here. There are 44 sites on the UK National Tide Gauge Network where this methodology could be extended. It would be interesting to apply our methodology within a spatial framework, for example in regional frequency analysis where sites in a homogeneous region not only have a common shape parameter, but also common longer-term trends due to anthropogenic climate change.

Skew surges are also believed to change over decadal time scales with climate indices. The North Atlantic Oscillation index (NAO) describes such time scale changes in regional weather systems, so is believed to impact storm surges, and thus skew surge.~\cite{AraujoPugh2008} find a negative correlation between storm surge and air pressure patterns, using NAO. It would be interesting to explore how adding an NAO covariate into the GPD for extreme skew surges would change model fit.

\section{Acknowledgments} 
This paper is based on work completed while Eleanor D'Arcy was part of the EPSRC funded STOR-i centre for doctoral training (EP/S022252/1). Thanks to Jenny Sansom of the Environment Agency and Joanne Williams of National Oceanography Centre for providing the data.

\bibliography{ref.bib}

\begin{thebibliography}{}

\bibitem[Ara{\'u}jo and Pugh, 2008]{AraujoPugh2008}
Ara{\'u}jo, I.~B. and Pugh, D.~T. (2008).
\newblock Sea levels at {N}ewlyn 1915--2005: analysis of trends for future
  flooding risks.
\newblock {\em Journal of Coastal Research}, (24):203--212.

\bibitem[Baranes et~al., 2020]{baranes2020}
Baranes, H., Woodruff, J., Talke, S., Kopp, R., Ray, R., and DeConto, R.
  (2020).
\newblock {Tidally driven interannual variation in extreme sea level
  frequencies in the Gulf of Maine}.
\newblock {\em Journal of Geophysical Research: Oceans}, 125(10):e2020JC016291.

\bibitem[Batstone et~al., 2013]{Batstone2013}
Batstone, C., Lawless, M., Tawn, J.~A., Horsburgh, K., Blackman, D., McMillan,
  A., Worth, D., Laeger, S., and Hunt, T. (2013).
\newblock {A UK best-practice approach for extreme sea-level analysis along
  complex topographic coastlines}.
\newblock {\em Ocean Engineering}, 71:28--39.

\bibitem[Coles, 2001]{Coles2001}
Coles, S.~G. (2001).
\newblock {\em An Introduction to Statistical Modeling of Extreme Values}.
\newblock Springer, London.

\bibitem[Coles et~al., 1999]{Coles1999}
Coles, S.~G., Heffernan, J., and Tawn, J.~A. (1999).
\newblock Dependence measures for extreme value analyses.
\newblock {\em Extremes}, 2(4):339--365.

\bibitem[D'Arcy et~al., 2022]{DArcy22}
D'Arcy, E., Tawn, J.~A., Joly, A., and Sifnioti, D.~E. (2022).
\newblock Accounting for seasonality in extreme sea level estimation.
\newblock {\em Under review}.
\newblock arXiv:2207.09870.

\bibitem[Davison et~al., 2012]{davison2012}
Davison, A.~C., Padoan, S.~A., and Ribatet, M. (2012).
\newblock Statistical modeling of spatial extremes.
\newblock {\em Statistical science}, 27(2):161--186.

\bibitem[Dixon and Tawn, 1999]{DixonTawn1999}
Dixon, M.~J. and Tawn, J.~A. (1999).
\newblock The effect of non-stationarity on extreme sea-level estimation.
\newblock {\em Journal of the Royal Statistical Society: Series C},
  48(2):135--151.

\bibitem[Dixon et~al., 1998]{dixontawnvassie1998}
Dixon, M.~J., Tawn, J.~A., and Vassie, J.~M. (1998).
\newblock Spatial modelling of extreme sea-levels.
\newblock {\em Environmetrics}, 9(3):283--301.

\bibitem[Eastoe and Tawn, 2009]{EastoeTawn2009}
Eastoe, E.~F. and Tawn, J.~A. (2009).
\newblock Modelling non‐stationary extremes with application to surface level
  ozone.
\newblock {\em Journal of the Royal Statistical Society: Series C},
  58(1):25--45.

\bibitem[Egbert and Ray, 2017]{egbert2017}
Egbert, G.~D. and Ray, R.~D. (2017).
\newblock Tidal prediction.
\newblock {\em Journal of Marine Research}, 75(3):189--237.

\bibitem[{Environment Agency}, 2018]{CFB2018}
{Environment Agency} (2018).
\newblock {Coastal Flood Boundary Conditions for the UK: update 2018. Technical
  summary report.}
\newblock
  \url{https://environment.data.gov.uk/dataset/6e856bda-0ca9-404f-93d7-566a2378a7a8}.
\newblock Accessed 01/10/21.

\bibitem[Fawcett and Walshaw, 2007]{FawcettWalshaw2007}
Fawcett, L. and Walshaw, D. (2007).
\newblock Improved estimation for temporally clustered extremes.
\newblock {\em Environmetrics}, 18(2):173--188.

\bibitem[Ferro and Segers, 2003]{FerroSegers2003}
Ferro, C. A.~T. and Segers, J. (2003).
\newblock Inference for clusters of extreme values.
\newblock {\em Journal of the Royal Statistical Society: Series B},
  65(2):545--556.

\bibitem[Graff, 1978]{graff1978concerning}
Graff, J. (1978).
\newblock Concerning the recurrence of abnormal sea levels.
\newblock {\em Coastal Engineering}, 2:177--187.

\bibitem[Howard and Williams, 2021]{howardWilliams2021}
Howard, T. and Williams, S. D.~P. (2021).
\newblock Towards using state-of-the-art climate models to help constrain
  estimates of unprecedented {UK} storm surges.
\newblock {\em Natural Hazards and Earth System Sciences}, 21(12):3693--3712.

\bibitem[Huser and Genton, 2016]{huser2016}
Huser, R. and Genton, M.~G. (2016).
\newblock Non-stationary dependence structures for spatial extremes.
\newblock {\em Journal of agricultural, biological, and environmental
  statistics}, 21(3):470--491.

\bibitem[Leadbetter et~al., 1983]{Leadbetter1983}
Leadbetter, M., Lindgren, G., and Rootz\'{e}n, H. (1983).
\newblock {\em {Extremes and Related Properties of Random Sequences and
  Processes}}.
\newblock Springer-Verlag, New York.

\bibitem[Ledford and Tawn, 2003]{LedfordTawn2003}
Ledford, A.~W. and Tawn, J.~A. (2003).
\newblock Diagnostics for dependence within time series extremes.
\newblock {\em Journal of the Royal Statistical Society: Series B},
  65(2):521--543.

\bibitem[Northrop et~al., 2017]{northrop2017}
Northrop, P.~J., Attalides, N., and Jonathan, P. (2017).
\newblock Cross-validatory extreme value threshold selection and uncertainty
  with application to ocean storm severity.
\newblock {\em Journal of the Royal Statistical Society: Series C (Applied
  Statistics)}, 66(1):93--120.

\bibitem[Pugh and Vassie, 1978]{PughVassie1978}
Pugh, D. and Vassie, J. (1978).
\newblock Extreme sea levels from tide and surge probability.
\newblock {\em Coastal Engineering}, 16:911--930.

\bibitem[Pugh and Woodworth, 2014]{pughwoodworth2014}
Pugh, D. and Woodworth, P. (2014).
\newblock {\em Sea-Level Science: Understanding Tides, Surges, Tsunamis and
  Mean Sea-Level Changes}.
\newblock Cambridge University Press.

\bibitem[Smith et~al., 1997]{SmithTawnColes1997}
Smith, R.~L., Tawn, J.~A., and Coles, S.~G. (1997).
\newblock Markov chain models for threshold exceedances.
\newblock {\em Biometrika}, 84(2):249--268.

\bibitem[Smith and Weissman, 1994]{SmithWeissman1994}
Smith, R.~L. and Weissman, I. (1994).
\newblock Estimating the extremal index.
\newblock {\em Journal of the Royal Statistical Society: Series B},
  56(3):515--528.

\bibitem[Southworth et~al., 2020]{texmex}
Southworth, H., Heffernan, J.~E., and Metcalfe, P.~D. (2020).
\newblock {\em texmex: Statistical modelling of extreme values}.
\newblock R package version 2.4.8.

\bibitem[Tawn, 1988]{tawn1988sealevels}
Tawn, J.~A. (1988).
\newblock An extreme-value theory model for dependent observations.
\newblock {\em Journal of Hydrology}, 101(1-4):227--250.

\bibitem[Tawn, 1992]{Tawn1992}
Tawn, J.~A. (1992).
\newblock Estimating probabilities of extreme sea-levels.
\newblock {\em Journal of the Royal Statistical Society: Series C},
  41(1):77--93.

\bibitem[Wadsworth and Tawn, 2012]{wadsworth2012}
Wadsworth, J. and Tawn, J. (2012).
\newblock Likelihood-based procedures for threshold diagnostics and uncertainty
  in extreme value modelling.
\newblock {\em Journal of the Royal Statistical Society: Series B (Statistical
  Methodology)}, 74(3):543--567.

\bibitem[Wahl et~al., 2013]{wahl2013}
Wahl, T., Haigh, I.~D., Woodworth, P.~L., Albrecht, F., Dillingh, D., Jensen,
  J., Nicholls, R.~J., Weisse, R., and W{\"o}ppelmann, G. (2013).
\newblock {Observed mean sea level changes around the North Sea coastline from
  1800 to present}.
\newblock {\em Earth-Science Reviews}, 124:51--67.

\bibitem[Weisse et~al., 2014]{weisse2014}
Weisse, R., Bellafiore, D., Men{\'e}ndez, M., M{\'e}ndez, F., Nicholls, R.~J.,
  Umgiesser, G., and Willems, P. (2014).
\newblock {Changing extreme sea levels along European coasts}.
\newblock {\em Coastal engineering}, 87:4--14.

\bibitem[Williams et~al., 2016]{Williams2016}
Williams, J., Horsburgh, K.~J., Williams, J.~A., and Proctor, R.~N. (2016).
\newblock Tide and skew surge independence: new insights for flood risk.
\newblock {\em Geophysical Research Letters}, 43(12):6410--6417.

\bibitem[Wong et~al., 2022]{wong22}
Wong, T.~E., Sheets, H., Torline, T., and Zhang, M. (2022).
\newblock Evidence for increasing frequency of extreme coastal sea levels.
\newblock {\em Frontiers in Climate}, 4.

\bibitem[Woodworth et~al., 2011]{woodworth2011}
Woodworth, P.~L., Men{\'e}ndez, M., and Roland~Gehrels, W. (2011).
\newblock Evidence for century-timescale acceleration in mean sea levels and
  for recent changes in extreme sea levels.
\newblock {\em Surveys in geophysics}, 32(4):603--618.

\bibitem[Woodworth and Player, 2003]{woodworth2003}
Woodworth, P.~L. and Player, R. (2003).
\newblock The permanent service for mean sea level: An update to the 21st
  century.
\newblock {\em Journal of Coastal Research}, 19(2):287--295.

\bibitem[Zsamboky et~al., 2011]{zsamboky2011}
Zsamboky, M., Fern{\'a}ndez-Bilbao, A., Smith, D., Knight, J., and Allan, J.
  (2011).
\newblock {Impacts of climate change on disadvantaged UK coastal communities}.
\newblock {\em Joseph Rowntree Foundation}, pages 1--63.

\end{thebibliography}
\bibliographystyle{apalike}

\end{document}